\documentclass[fleqn,12pt,twoside]{article}
\usepackage{espcrc1}

\usepackage{graphicx}
\usepackage[figuresright]{rotating}

\newcommand{\AmS}{{\protect\the\textfont2
  A\kern-.1667em\lower.5ex\hbox{M}\kern-.125emS}}

\title{X-ray binaries}

\author{
H. SCHATZ
\address{National Superconducting Cyclotron Laboratory\\
Dept. of Physics and Astronomy\\
Joint Institute for Nuclear Astrophysics\\
Michigan State University\\
East Lansing, MI 48824}
and
K. E. REHM
\address{Physics Division\\
Argonne National Laboratory}}

\begin{document}

\maketitle

\section{Introduction}

X-ray binaries are among the brightest extra-solar objects in the sky
and are characterized by dramatic variabilities in brightness
on timescales ranging from milliseconds to months and years.
Their main source of power is the gravitational
energy released by matter accreted from
a companion star and falling onto 
a neutron star or a black hole in a close binary system. X-ray binaries 
therefore serve as rich sources of information about 
compact stellar objects, and, once understood, could be 
used as unique natural laboratories for the properties of 
matter under extreme conditions. 

In recent years, the launching of a number of X-ray 
observatories has marked the beginning of 
a new era in X-ray astronomy. These
include Beppo SAX, RXTE, Chandra X-ray observatory and
XMM-Newton. These facilities provide(d) unprecedented 
sensitivity, all-sky coverage, and timing 
resolution and led to a range of new discoveries related
to X-ray binaries, such as ms-oscillations, superbursts, and 
quiescent luminosity measurements in transients.
Similar advances in the underlying nuclear physics are now needed to make 
full use of these observations towards a better 
understanding of the physics of X-ray binaries, and 
to solve the many open questions. 
In this review we concentrate on accreting neutron star 
systems, where observables are to a large extent governed
by nuclear physics. For some recent reviews see
Psaltis 2004 \cite{Psa04} or Strohmayer \& Bildsten 2003 \cite{StB03}.

In section \ref{SecNucObs} we summarize recent observational
results, open questions, and their relation to the underlying nuclear physics.
In section \ref{SecNuc} we summarize the current 
status of the nuclear physics, identify the major uncertainties,
and give an outlook on the role that future radioactive
beam facilities such as RIA can play in this field. 

\section{Nuclear processes and their observables}
\label{SecNucObs}

A fluid element consisting of hydrogen and helium that is accreted
onto the surface of a neutron star is continuously 
compressed by the matter accumulating on top of it. Eventually 
the element will become part of the liquid ocean and later of 
the solid crust of the neutron star. 
The thin outer crust of a neutron star represents just
$\approx 10^{-4} M_\odot$ of material. At typical accretion 
rates of 10$^{-8}$-10$^{-10} M_\odot/{\rm yr}$
this material is therefore replaced after 
about 10$^{4}$-10$^6$ years of accretion. This is short
compared to the duration of 
the mass transfer phase in low mass X-ray binaries. 
Therefore the majority of neutron stars in low mass X-ray binaries
should have an accreted crust. 

On its journey into the neutron star interior the accreted fluid
element undergoes a sequence of nuclear processes converting
hydrogen and helium into 
exotic nuclei ranging from the proton drip line during 
X-ray bursts to the neutron 
drip line upon entering the inner crust. 
These processes together with their relation to 
observations and current open questions are discussed in the 
following subsections in the order they occur. The ashes of 
each process form the initial composition for the next, and 
the energy release of a nuclear process deep in the crust
sets the thermal environment for the surface processes. Therefore,
all processes are connected and need to be understood in 
a selfconsistent way. 

\subsection{Spectral features and spallation of metals}

During the 10-100 s of an 
X-ray burst (see Sec. \ref{SecXRB}) the X-ray flux is dominated by 
thermal emission from the dramatically heated neutron star 
photosphere. This offers the opportunity to obtain 
compositional information from absorption features in the 
X-ray spectrum. In addition, observed spectral features
allow one to determine the 
gravitational redshift on the neutron star surface 
and therefore the compactness (mass/radius) of the neutron star. 

Some observations of a $\approx$ 4~keV
absorption feature have been reported over the last 20 years for
various systems (see recent review
in Bildsten, Chang, and Paerels 2003 \cite{BCP03}). 
More recently, Cottam, Paerels, and Mendez 2002 \cite{CPM02}, reported the 
detection of hydrogen and helium like Fe absorption lines and
derive a redshift of $Z=0.35$ for the neutron star in EXO 0748-676.
This has provided an important constraint for the neutron star 
mass radius relation and the nuclear equation of 
state. In addition
a recently discovered 45 Hz oscillation during an X-ray burst
\cite{StV04} in EXO 0748-676 has been associated with the 
spin frequency of the underlying neutron star (see Sec.~\ref{SecXRB}). 
The neutron star in EXO 0748-676 apparently rotates much slower
than other neutron stars in X-ray bursters that typically spin
with frequencies ranging from 270 to 620~Hz \cite{StB03}. 
The reduced rotational line broadening in EXO 0748-676 might 
be the reason why a spectral line could be detected in this system.
The known spin frequency also offers
the opportunity to use precise modeling of the spectral line shapes
to constrain the neutron star radius
and therefore the dense matter equation of state.

As Joss 1977 \cite{Jos77} pointed out, 
the large entropy difference between the deeper layers that 
undergo explosive hydrogen and 
helium burning during X-ray bursts (\ref{SecXRB}) and the photosphere
prevents in principle convection from transporting burst ashes to the surface
(see however Weinberg, Bildsten, and Schatz 2006 \cite{Nevin06} 
for exceptions related to photospheric
radius expansion bursts).
On the other hand, heavy elements such as Fe entering the 
photosphere via accretion, sink within 
seconds beneath the photosphere owing to the strong gravity on 
the neutron star surface \cite{BCP03}. Elemental abundance observations in
the neutron star photosphere 
during X-ray bursts therefore probe directly the ongoing 
accretion and provide information 
on the accreted composition as well as the trajectories of the 
infalling matter. In particular, 
infall under steep angles can lead
to the spallation of heavy elements \cite{BSW92} contained in 
the accretion flow.  This has been 
investigated in detail by Bildsten, Chang, 
and Paerels 2003 \cite{BCP03}, who follow the cascade of fragmentation 
processes and predict the final abundances. They find, that 
the amount of iron that can be present in the photosphere is about a factor of 20 
short of explaining the observed line strength, but argue for 
more accurate studies before drawing definite conclusions. 

Spallation of heavy elements is also important for X-ray burst 
calculations. The abundance
of CNO elements entering the deeper atmosphere of the
neutron star controls how much hydrogen can be burned away 
by the CNO cycle prior to the triggering of an X-ray burst.

\subsection{X-ray bursts and the rp-process} \label{SecXRB}

Shortly after the discovery of X-ray bursts in 1976 \cite{Gri76,BCE76}
they were correctly interpreted as thermonuclear explosions
on the surface of an accreting neutron star \cite{WoT76,MaC77,Jos77,LaL78}. 
Hydrogen or helium-rich matter from 
the envelope of the companion star accumulates on the neutron star
for hours to days. Eventually the ignition of extremely 
temperature sensitive
fusion reactions leads to the thin shell instability of Hansen 
and VanHorn 1975 \cite{HaV75}, where the heating from thermonuclear reactions 
cannot be compensated by readjusting the stellar structure or 
cooling through the surface. The 
result is a thermonuclear runaway where rising temperatures 
accelerate the thermonuclear reaction rates leading to more
rapid temperature rises, still faster thermonuclear reactions
and so on. The resulting thermonuclear
energy release lasts 10-100 s and can be directly observed 
as an X-ray outburst of typically 10$^{39-40}$ ergs. 
Burst recurrence times are of the order of hours to days and 
long sequences of burst can be observed. About 60 galactic X-ray 
bursting systems are known today making X-ray bursts 
a frequent phenomenon in our Galaxy (see \cite{Psa04,Bil97,LPT93} for 
reviews on observations and physics of X-ray bursts). 

Over the last years a dramatic increase in observational data 
has led to a large database of burst properties monitored
over long periods of time, and has also led to new discoveries
such as ms oscillations of the X-ray flux during X-ray bursts
\cite{Str96,StB03}.
These oscillations have been interpreted as 
anisotropies in the nuclear burning on the surface of the 
rapidly rotating neutron star, caused for example by a 
spreading hot spot. In this model the slight frequency changes observed
over the burst duration are explained by the rotational decoupling of 
the surface layer that slows down
during the burst driven expansion of the neutron star atmosphere
and then reaccelerates during the following cooling and contraction. 
However, in some cases 
the frequency changes of the oscillations
are too large to be explained with a rotationally
decoupled layer \cite{CMB02}. Alternative explanations for
the frequency drifts include various surface oscillation 
modes \cite{PiB05,CuB98,MVH88}. The frequencies of these modes
depend on the composition and therefore on the 
nuclear physics of X-ray bursts. 
In any case, the recent observations of ms oscillations during X-ray bursts
in the two known X-ray bursting pulsars, SAX J1808.4-3658 
\cite{CMM03} and XTE J1814-338 \cite{SMS03}, where 
the spin frequency is known from the regular pulsations, 
have finally proven the relationship between burst oscillations
and neutron star spin. 

Observations of ms oscillations are important for 
many reasons. First they allow to experimentally explore 
the X-ray burst behavior and the underlying nuclear processes
as a function of neutron star spin
\cite{MGC04}. Second, as Strohmayer 2004 \cite{Str04} pointed out
gravitational light bending effects in principle would allow one to extract 
mass and radius of the neutron star together with the time-dependent 
size of the surface burning area from a detailed analysis of pulse
trains. Already first constraints on 
the mass radius relationship of the neutron star in XTE J1814-338
have been obtained with this method \cite{BSC04}.

The nuclear processes powering X-ray bursts and the related 
observational features
depend strongly on 
the system parameters, such as mass accretion rate, 
accreted abundances of hydrogen, helium, and CNO metals, rotation,
and the thermal flux at the surface originating from 
nuclear reactions deep in the crust \cite{Bil97}. 

Most systems have mass accretion rates in excess of 
4.4$\times 10^{-10}
M_\odot/yr$
and accrete a mix of hydrogen, helium and some heavy metals 
most likely 
in solar proportions. In this case, the burst is triggered by 
the temperature 
sensitivity of helium burning via the 3$\alpha$ reaction, igniting 
helium in a hydrogen rich surface layer \cite{Bil97}.
The helium ignition triggers hydrogen burning via the 
rapid proton capture process (rp-process) \cite{WaW81,Van94,SAG97}
and peak temperatures of 1-2~GK are reached. 

Because of the factor
of 10 larger energy release per nucleon, hydrogen 
burning via the rp-process is the main energy source 
for X-ray bursts and determines the X-ray light curve. 
Nevertheless, because of its computational demanding 
nature, the full rp-process has only been recently 
included in X-ray burst calculations, opening the 
door for quantitative comparisons with observations. 
In a first step, Schatz et al. 2001 \cite{SAB01} included the 
full rp-process in a one-zone model after exploratory studies
with smaller networks \cite{SAG97,WiS99}.
This led to the discovery of a SnSbTe cycle which 
is formed by ($\gamma$,$\alpha$) reactions on $\alpha$-unbound 
proton-rich Te isotopes.
The resulting rp-process path is shown in Fig.~\ref{FigRp} 
for an X-ray burst where
large amounts of hydrogen are available at burst ignition. 
The SnSbTe cycle represents a natural end point for any
single outburst rp-process. 
In principle heavier elements could be produced in a 
multi-burst rp-process where the heavy ashes decay back to stability
and are then again irradiated
with protons thereby bypassing 
the Sn-Sb-Te cycle \cite{BHM98}. So far there is no plausible
astrophysical scenario known where this would occur. 

Besides delineating the full rp-process path for the first time,
three main conclusions were drawn from 
these one-zone calculations.
First, hydrogen is always burned completely
making the occurrence of deep hydrogen burning via electron
capture \cite{TWL96} unlikely. Second, the rp-process
beyond Fe includes particularly long-lived $\beta$-decays
at $^{64}$Ge, $^{68}$Se, $^{72}$Kr and $^{104}$Sn
(so-called waiting points) that
slow the process down and extend the energy release of
the burst considerably. The result is an increase of
the burst duration from a few 10~s of seconds to 100-300 s.
Qualitatively similar conclusions have been drawn by
Koike et al. 1999 \cite{KHA99} based
on a similar one-zone model but with a more limited reaction
network. As a consequence, the observation of long X-ray bursts
with durations of the order of 100~s can now be used as an indicator
for mixed hydrogen and helium burning providing a powerful constraint
for models and system parameters.
The third conclusion drawn from one-zone X-ray burst calculations
is that the  long-lived
waiting points along the rp-process path lead
to a wide spread of the final abundance
distribution \cite{SAB01,SBC98}. Therefore the crust of 
an accreting neutron star is characterized by a 
very impure lattice structure that strongly affects crust 
properties such as thermal and electrical conductivities
(see Sec. \ref{SecCrust}). 
Koike et al. 2004 \cite{KHK04} obtain similar results by combining 
a full one-dimensional X-ray burst multizone model with a very limited 
nuclear reaction network 
to calculate temperature and density conditions followed
by postprocessing with a full reaction network. 
However, it has been shown that even at late stages of the burst
nuclear reactions on heavier nuclei beyond iron 
produce significant amounts of energy and can still determine 
temperature and observed light curves
\cite{KHA99,SAB01}. Therefore, for X-ray bursts postprocessing 
does not reproduce the correct conditions and nuclear reaction 
flows (see also \cite{TBF01} for an example). 

More recently the first one-dimensional multizone calculations 
of a sequence of X-ray bursts have been performed with the full rp-process
\cite{WHC04,FTW04}. 
These calculations confirmed the main conclusions from the earlier 
one-zone models, but also yielded some important differences. Most 
importantly, the ignition conditions for bursts late in the 
burst sequence change because of the presence of partially 
burned ashes from preceding bursts.  
This is the well known compositional inertia effect already 
pointed out by Taam (1980) \cite{Tam80}. The main consequence is increased hydrogen 
burning between bursts leaving less hydrogen for the 
X-ray bursts itself. Nevertheless, bursts at high accretion rates still 
burn enough hydrogen for the rp-process to reach the first
long-lived waiting points in the $A=64$ region and therefore
still lead 
to long duration X-ray bursts. For the few system 
parameters explored so far the 
SnSbTe cycle is only reached in the very first burst after 
the onset of accretion. This is of potential interest 
for transient systems, where the observation of the 
sequence of bursts after the accretion turns on could 
provide important constraints for X-ray burst models
and the structure of the underlying neutron star. 
In the system Acq X-1 such a first burst might have been 
observed \cite{CCG87}, and its unusual long duration would 
be consistent with a particularly extended phase 
of hydrogen burning \cite{FTW92}

A first quantitative comparison of calculated burst profiles
from the one-dimensional model of Woosley et al. \cite{WHC04}
with the so-called "textbook" burster GS 1826-24 has yielded
excellent agreement \cite{Cum03} for certain 
choices in the nuclear physics. 
GS 1826-24 exhibits extremely regular bursts with a period of 3-6 hours
and observation
over many years has yielded very accurate burst profiles.
Galloway et al. 2004 \cite{GCK04} in fact find a slight change in burst profile
from observations in the years 1998 and 1999 to observations
in the year 2000 due to a slight increase in accretion rate. 
It remains to be seen whether these changes can be reproduced
in current models, but observations like this certainly 
provide unprecedented quantitative tests for X-ray burst 
models. However, the nuclear physics of the rp-process is far 
from being sufficiently accurate to test X-ray burst models
at this level.
This has been clearly demonstrated with studies of 
the sensitivity of calculated X-ray burst light curves 
to uncertainties in masses around the major waiting points $^{64}$Ge, 
$^{68}$Se, and $^{72}$Kr in a one-zone model
\cite{AB1}. More recent studies with multi-zone X-ray 
burst models predicting light curves for various 
changes in $\beta$-decay half-lives, which were intended
to simulate mass uncertainties, come to similar
conclusions \cite{WHC04}. 
It is likely that these studies rather underestimate
the nuclear physics uncertainties in predicted X-ray burst 
light curves as there are additional, large uncertainties 
in nuclear reaction rates (see Sec.~\ref{SecNuc}). 
Clearly, the current status of the nuclear physics of the 
rp-process does not allow full interpretation of X-ray observations
in a quantitative way. The success of modeling the 
burst profiles of GS 1826-24 is promising but does therefore
not necessarily imply that current astrophysical models
and their parameter choices are correct. 

There are many other open questions related to the nuclear 
burning in X-ray bursts.
For example,
the ignition and subsequent burning front propagation across
the neutron star surface is not understood. This 
relates directly to the observations of ms oscillations
during bursts which could originate from
unisotropic burning across the rapidly spinning neutron star surface 
when the accreted layer ignites at one, or a few
spots or patches. 
First attempts with analytical 
considerations or simplified models have been 
made to investigate the spreading of a burning 
front across the neutron star surface 
\cite{FrW82,Bil95,SLU02}. 
They find that the burning front is 
most likely a deflagration front driven by 
convection. As Zingale et al. \cite{Zin02}
point out a localized
ignition does not seem possible,
at least when neglecting rotation.
It would be desirable to 
explore this in full 2D or 3D hydrodynamic 
calculations. Such calculations have so 
far only been done for the first 150~$\mu$s
in the much easier case of a
pure He detonation \cite{Zin01}. 

Another major problem in understanding X-ray bursts is the 
dramatic change in bursting behavior once the 
accretion rate exceeds about 0.13 $\dot{M}_{\rm Edd}$.  
Current simple X-ray 
burst models predict that with increasing 
accretion rate ignition condition
are reached earlier leading to more frequent bursts. 
At the same time the burst duration should become longer as
there is less time to burn hydrogen during the accretion phase
leaving more fuel for the slow rp-process in bursts. 
While this behavior is qualitatively confirmed for the 
slight change of accretion rate in the textbook 
burster GS 1826-24 \cite{WHC04}, the observed behavior 
beyond 0.13 $\dot{M}_{\rm Edd}$
is exactly opposite \cite{Cor03}. 
This can be seen very clearly for KS 1731-260 where 
burst observations exist for accretion 
rates ranging from 0.04 to 0.4 $\dot{M}_{\rm Edd}$ \cite{Cor03}. 
Beyond an accretion rate of 0.13 $\dot{M}_{\rm Edd}$
bursts suddenly become rarer and 
shorter, indicating that there is not 
enough hydrogen for the rp-process to reach the 
first major waiting point at $A=64$.
Furthermore, the so-called $\alpha$ value, 
the ratio of energy released in between bursts to the energy 
released in bursts, suddenly increases from 40
(expected if all nuclear burning happens in bursts)
to 500-5000. This indicates that beyond 0.13 $\dot{M}_{\rm Edd}$
most of the nuclear burning occurs outside of bursts in some 
stable fashion, though current spherically symmetric
X-ray burst models predict that all fuel should be burned
in bursts.
Bildsten 1995 proposed that some nuclear burning 
proceeds through a slow deflagration front moving across the 
neutron star surface instead of a rapidly burning burst \cite{Bil95}.
Another 
explanation is given by Narayan and Heyl 2003 \cite{NaH03}, who 
performed a linear stability analysis of 
the accreted envelope and find a new 
instability regime of so-called "delayed-mixed bursts"
at accretion rates between 
0.14 and 0.25 $\dot{M}_{\rm Edd}$,
where stable burning and X-ray bursts coexist.
The predicted accretion rate for the transition into this 
new regime agrees
very well with observations. However, the model 
also predicts a transition to stable burning 
and the disappearance of bursts 
beyond an accretion rate of 
0.25 $\dot{M}_{\rm Edd}$ while there are 
bursts observed at accretion rates as
high as 0.7 $\dot{M}_{\rm Edd}$ \cite{Cor03}. 
Furthermore, the "delayed-mixed bursting" behavior (though observed in nature)
is not found in recent time-dependent 
multizone calculations \cite{Heg04}
which should in principle give the same result.
The lack of understanding of X-ray bursts at accretion 
rates beyond 0.13 $\dot{M}_{\rm Edd}$
is particularly bothersome as this 
is exactly the regime relevant for superbursts (see below). 

Another long-standing open question is whether 
X-ray bursts do in some form contribute to Galactic 
nucleosynthesis \cite{HBM95,SAG97}. This would be interesting as the 
rp-process can produce significant amounts 
of $^{92,94}$Mo and $^{96,98}$Ru. These p-nuclei 
cannot be produced in sufficient amounts in 
standard p-process models \cite{LAM92,ArG03}.
Costa et al. 2000 \cite{CRZ00} proposed that a much larger
than theoretically expected $^{22}$Ne($\alpha$,n)
reaction rate in the s-process, which produces
the p-process seeds, could solve the problem. 
However, theoretical arguments and recent 
experimental upper limits seem to rule out
such a large $^{22}$Ne($\alpha$,n) reaction 
rate \cite{JKM01}. The production
of proton-rich Mo and Ru isotopes 
at another site
would also solve the problem of 
the origin of these isotopes in the solar system. 
One alternative site that has been proposed is a
combination of a weak rp-process with
neutron captures in explosions of white dwarfs 
that reach the Chandrasekhar mass limit by 
accreting a helium layer \cite{GJH02}. 
X-ray bursts could in principle be another 
possibility.

However, the critical question is how matter containing
rp-process ashes can be ejected from the surface of 
a neutron star. 
The fundamental problem is to overcome the 
gravitational binding of 200 MeV per nucleon 
when the nuclear energy release cannot exceed 
6 MeV per nucleon. Recently it has been shown that
the convective zone developing during the rise of 
an X-ray burst can extend to sufficiently shallow
regions that can be blown off in winds. Such winds are thought to 
develop during the frequently observed and particularly 
luminous photospheric expansion bursts \cite{Nevin06}. This provides
a mechanism to eject ashes from thermonuclear burning 
into space in a subclass of X-ray bursts. An 
observational confirmation, for example through detection 
of spectral features imprinted onto the X-ray flux
by the ejected material would be extremely important. 
However, the material transported to the surface 
is the ashes of the initial stages of the 
thermonuclear burning and even for hydrogen 
rich bursts is dominated by products of the $\alpha p$ 
process, mainly $^{24}$Mg and $^{28}$Si, with some
smaller amounts of nuclei up to $A \approx 38$.
Small amounts of $^{60,62}$Zn from 
processing of the accreted iron are also mixed in. 
Therefore, this 
still does not necessarily provide a mechanism
to eject heavy rp-process ashes produced
during the cooling phase of the X-ray burst when 
convection has stopped. However, in principle
ashes mixed into the burning zone of subsequent 
bursts or mass ejection through a superburst
might be possibilities \cite{Nevin06}. 

Another problem is the total amount of Mo and Ru
that can be produced in the Galaxy by X-ray bursts. 
While Schatz et al. 1998 \cite{SAG97} pointed out that in principle
with the ejection of just 1\% of the processed matter 
X-ray bursts could be contributors to Galactic nucleosynthesis,
new calculations show \cite{WHC04}
that for the rp-process to produce Mo and Ru 
special circumstances are needed. In the 
majority of X-ray bursts the rp-process will not produce
significant amounts of Mo and Ru. Furthermore, the existence
of the later discovered SnSbTe cycle \cite{SAB01} leads to 
a large co-production of $^{102}$Pd and $^{106}$Cd whenever the rp-process
reaches the $A \approx 90-100$ mass region. 
$^{102}$Pd and $^{106}$Cd do not share the unusual isotopic enhancement 
found for most proton-rich Mo and Ru isotopes and are produced in sufficient 
amounts in standard p-process scenarios. 
It has also been pointed out that the long-lived radioactive
p-nucleus $^{92}$Nb might provide important clues 
as it is co-produced with $^{92}$Mo and 
other p-nuclei in a photodisintegration type 
supernova p-process, but is shielded against 
rp-process contributions by stable $^{92}$Mo \cite{DRM02}. Indeed, 
traces of extinct $^{92}$Nb present in the early solar system 
have been detected in primitive meteorites. 
Estimates using the measured abundance and a galactic chemical evolution model
find a $^{92}$Mo/$^{92}$Nb production ratio that is very close
to the one predicted from supernovae p-process models. 
However, there are large uncertainties in the galactic
chemical evolution modeling and the nuclear physics that determines
the $^{92}$Mo/$^{92}$Nb ratio in supernovae. 
In summary it seems to be difficult to explain
the origin of proton-rich Mo and
Ru isotopes with the 
rp-process in X-ray bursts based on current models. 
Nevertheless, the question of the 
ejection of burned material in X-ray bursts and its composition 
should be explored further
not only in terms of the nucleosynthesis contribution but also 
to explore the possibility of future abundance 
observations through X-ray spectroscopy. 

\subsection{Superbursts and deep carbon burning}

Superbursts are rare, extremely powerful X-ray bursts
that have been discovered trough long term monitoring with the
Beppo-SAX Wide Field Camera and the RXTE 
All-Sky Monitor. They are found in X-ray binary systems that 
otherwise show regular X-ray bursts.
So far, 13 superbursts 
have been detected from 8 sources, most recently 
in GX $17+2$ \cite{ICC04} (see \cite{ICK04} for a recent 
observational overview).
Compared to the regular 
bursts, superbursts are a factor of 1000 times 
more energetic, they last many hours as opposed to 
10-100s, and their recurrence time is around a year
instead of hours to days. Superbursts show the 
same spectral behavior than normal X-ray bursts, 
and in one case it was 
possible to detect ms oscillations with RXTE
that show a behavior similar to the oscillations in 
normal bursts \cite{StM02}. Superbursts are therefore likely to be 
of thermonuclear origin. 

Superbursts most likely occur when 
carbon in the ashes from regular bursts ignites deep
in the neutron star's liquid surface ocean
\cite{WoT76,CuB01,StB02}. Cumming and Bildsten 
2001 \cite{CuB01} pointed out that the presence of 
heavy nuclei beyond iron in the ashes plays 
an important role as they increase the opacity 
of the layer and therefore affect the ignition depth. 
Schatz et al. 2003 \cite{SBC03} investigated in more detail the 
actual nuclear processes occurring during a superburst
and found that in addition to carbon fusion, photodisintegration
of heavy elements plays an important role. Once 
carbon burning heats the layer to 1-2~GK (depending on 
the detailed composition) the extremely temperature 
sensitive photodisintegration reactions on heavy 
nuclei trigger a photodisintegration runaway that leads
to a rapid rise of the temperature up to 7~GK. This drives 
the composition into Nuclear Statistical Equilibrium (NSE),
which at superburst conditions contains mainly nuclei around 
$^{66}$Ni.
Depending on the initial composition set by the ashes of the 
regular bursts the photodisintegration of heavy elements into 
Ni can contribute up to about 50\% of the total superburst energy. 

For ashes typical of explosive hydrogen burning
carbon ignition occurs around a density of 
10$^9$~g/cm$^3$ 
- about 3 orders of magnitude larger
than for ordinary X-ray bursts. The much higher ignition depth 
explains the longer recurrence time and the higher total energy
as more matter needs to be accumulated to trigger the burst.
The longer burst duration matches the radiative cooling timescale
from the ignition layer. The carbon ignition model also explains
the observed interaction of superbursts with normal bursts. 
The increasing heat flux at the surface shortly after superburst 
ignition triggers the explosive burning of the hydrogen and helium 
in the upper layers leading to the observed "precursors" - a regular
X-ray bursts preceding the superburst \cite{CuM04}. The long cooling timescales
lead to an increased surface heat flux for many weeks explaining 
the suppression of regular bursts after a superburst. 

Superbursts are an important probe for the properties of 
neutron stars. As they occur deeper than regular X-ray bursts, 
they are more sensitive to the thermal properties of the crust 
and core cooling models. As Brown \cite{Bro04} showed recently, 
the mere existence of superbursts can put tight constraints on 
the thermal conductivity of the neutron star crust and on 
the rate of neutrino cooling in the core. 

The main open question concerning superbursts is the origin of 
a sufficient amount of carbon (at least a few \% mass fraction) in the ashes
of hydrogen and helium burning on the surface. Recent one-zone 
models of X-ray bursts showed that when one 
includes the full rp-process hydrogen is consumed completely
leading to a brief helium burning phase at the end of the burst 
that produces some carbon. 
However, more recent calculations \cite{WHC04}
that follow a series of bursts, demonstrated that any carbon 
leftover in the ashes of a specific burst is destroyed via the 
capture of residual helium
triggered by the heat from the next burst. 
A key issue in this problem is the unexplained regular burst behavior
at accretion rates beyond 0.13 $\dot{M}_{\rm Edd}$, which is
just the range relevant
for superbursts (see Sec. \ref{SecXRB}). As discussed above, 
at these accretion rates some fuel is likely burned steadily 
and not in bursts. Such stable burning could in principle 
produce the amount of carbon needed to explain 
the superburst phenomenon \cite{SBC98}. 

Except for the carbon fusion rate that triggers the unstable burning
uncertainties in the nuclear physics of the processes during 
superbursts are not likely to be important. The matter is quickly 
driven into NSE, where final composition and
energy generation are determined by ground state masses and 
partition functions of experimentally well studied nuclei 
near stability. However, the nuclear physics
uncertainties in regular X-ray bursts, in particular in the rp-process,
together with the astrophysical puzzles associated with them 
prevent a reliable
calculation of the X-ray burst ashes which serve as the fuel
for superbursts. This is a major obstacle
towards a full understanding of superbursts and a solution of this 
problem is critical for using superbursts as reliable probes
for the properties of accreting neutron stars. 

\subsection{Crust processes} \label{SecCrust}

The neutron star crust serves as an interface between
the physics of the neutron star and observations. One 
direct observable
is the quiescent luminosity from the crust of neutron stars in 
transient X-ray binary systems (so-called Soft X-ray Transients). These 
systems 
are characterized by a thermal accretion disc instability that turns the 
accretion onto the neutron star on and off with periods ranging from
months to decades \cite{CSL97,CCM98}. 
It has been argued \cite{BBR98,RBB00} that the quiescent luminosity 
that is observed from Soft X-ray 
Transients after the accretion has turned off, is 
the thermal radiation from the cooling neutron star crust. 
This offers a pathway to distinguish neutron star systems from 
black hole systems, and to determine neutron star radii provided
the observed spectrum can be accurately modeled \cite{RBB99} and 
the distance to the source is known. 

In addition, 
luminosity measurements directly constrain the thermal 
state of the crust, which in turn can be used as a probe for
neutron star physics such as the existence of enhanced neutrino cooling processes,
superfluidity, or exotic phases of nuclear matter \cite{CGP01,YLH03}.
Wijnands 2004 \cite{Wij04} gives an extensive review of the observational status
concerning the detection of soft X-ray transients in quiescence. 
The majority of X-ray
transients have periods of the order of months and 
crust luminosity measurements 
can be used to determine the limit cycle thermal state of the crust
\cite{BBR98}.
In long period transients (so-called quasi-persistent sources),
which turn off for years to decades
repeated observations during the off state offer the unique opportunity to measure the crust 
cooling behavior of an accreting neutron star as a function of time \cite{RBB02}.
This has recently become possible in two cases - for KS 1731-260, which 
turned off early in 2001 a drop 
of the surface luminosity was followed in 4 measurements over 2.5 years
\cite{Wij04}. Similarly, for MXB 1659-29 3 measurements over several years 
were performed \cite{Wij04a}. In both cases the decrease in luminosity
of about a factor of 8 over a few years is consistent with the observed
decrease in effective temperature further supporting the assumption 
that one observes a cooling neutron star crust. Interestingly, 
comparison with crust models shows that such a rapid drop in 
luminosity can only be reproduced when assuming 
enhanced neutrino cooling in the neutron 
star core.
However, the detection of a superburst from KS 1731-260 during 
its on-state points to a hot crust and seems to rule out 
enhanced neutrino cooling \cite{Bro04}. At this point a solution to this puzzle
is not in sight, but clearly a better understanding of the nuclear
physics underlying both probes - superbursts and the neutron star 
surface luminosity behavior - would be important. 

The neutron star crust composition and its thermal state 
also affect the crust electrical conductivity. This 
is important for another fundamental open question - 
the existence of two classes of X-ray binaries: 
X-ray pulsars characterized by high magnetic fields that 
funnel the accretion flow onto the polar caps, and 
X-ray bursters with low magnetic fields. While a 
long term decay of magnetic fields seems to be a likely 
explanation, neither the origin of the magnetic fields
nor the mechanism of the field decay are understood
\cite{UGK98,BrB98,CAZ04}. 

As pointed out above, the initial composition of the neutron 
star crust is determined by the ashes of the surface 
burning processes such as X-ray bursts and superbursts.
The ongoing accretion compresses the ashes continuously
and converts it into deeper and deeper layers of the 
crust. The rising electron Fermi energy initiates a
sequence of electron captures, which beyond neutron 
drip are accompanied by neutron emission. In the inner 
crust of the neutron star the nuclei have finally 
low enough charge numbers to enable 
pycnonuclear fusion reactions
\cite{HaZ90,HaZ90a,HaZ03}. All these reactions generate heat and 
might also deform the crust leading to the possibility 
of gravitational wave emission if the neutron star 
is spinning rapidly \cite{Bil98,UCB00}. 

Together with the rp-process and the nuclear reactions
during superbursts these crust processes determine 
the crustal heating and the crust composition, which 
in turn strongly affects thermal and electrical 
crust conductivities. Therefore, the nuclear physics
of a wide range of processes  
needs to be understood
before observations can be translated into quantitative
constraints of neutron star properties. These processes
include all nuclei from the proton drip line to the neutron drip
line up to a maximum mass number that is set by the endpoint 
of the rp-process, with a maximum at $A\approx 106$ set by 
the Sn-Sb-Te cycle. 

So far, the nuclear processes in the crust of an accreting neutron
star have only been explored in a schematic way 
assuming single nuclide compositions, zero temperature and infinitely fast electron 
capture at threshold. While these are not unreasonable assumptions
for a first exploration of nuclear crust processes, more detailed simulations
would be needed to determine the impact of nuclear physics
uncertainties such as
electron capture thresholds, the location of the neutron drip line, 
electron capture rates, and pycnonuclear fusion rates on 
crust studies and observables.
In addition, crust processes strongly depend on the initial
composition set by the ashes of X-ray bursts and superbursts. 
The nuclear physics uncertainties in the rp-process therefore
strongly affect the physics of the crust. Already, 
the inclusion of reactions beyond Ni in the rp-process
has had a significant impact in demonstrating that 
the crust consists most likely not of a pure single
nuclide lattice, for example Fe, but of a wider
range of nuclear species. This is a consequence of 
the long waiting points in the rp-process beyond $^{56}$Ni 
that lead to a spreading in the final abundance distribution. 
First estimates show that such an impure crust 
has a much reduced thermal and electrical conductivity
\cite{BrB98,Jon04}.

\section{Nuclear physics needs}
\label{SecNuc}

The surface of accreting neutron stars is characterized by a wide 
range of nuclear processes ranging from fragmentation at infall, 
over thermonuclear burning in X-ray bursts and superbursts, to 
electron captures and pycnonuclear fusion reactions in the neutron
star crust. These processes involve the majority of the nuclei between
the proton and the neutron drip line up to a maximum mass number set 
by the endpoint of the rp-process. In this picture, the rp-process during X-ray bursts
(and maybe during stable hydrogen burning as well) plays a central 
role as it directly determines one of the key observables, the
observed X-ray bursts, and because it sets the initial composition for 
all the deeper processes. At the same time, the
nuclear physics and its
impact on observables has been explored for the rp-process to a much 
greater extent than for any of the other processes on 
accreting neutron stars. We therefore concentrate 
in the following on the open nuclear physics questions related to 
the rp-process. Earlier reviews of the experimental and theoretical 
nuclear physics aspects of the rp-process
can be found in \cite{ChW92,Van94,SAG97,WSC98,WiS99}.

\begin{figure}
\includegraphics[width=8in]{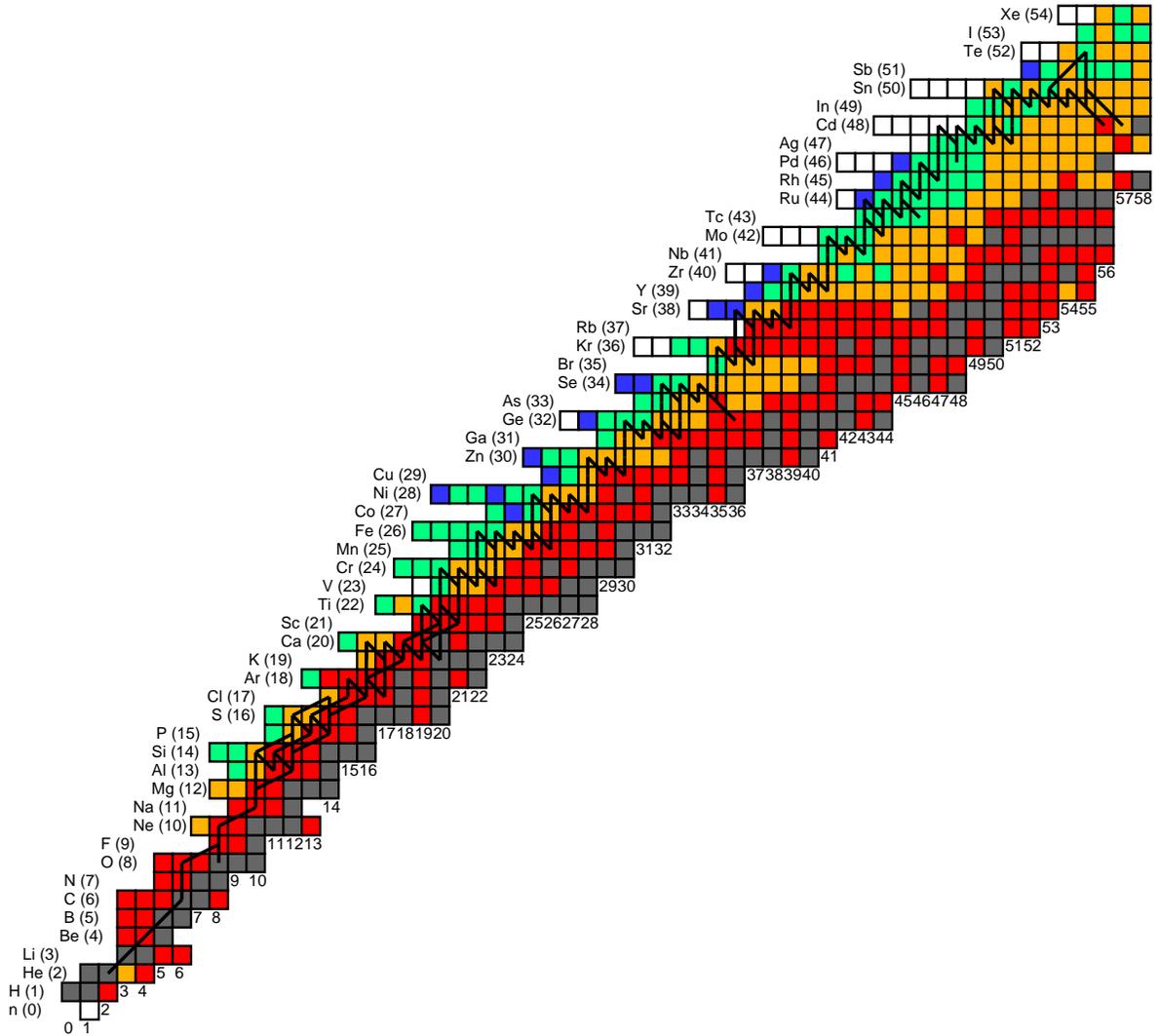}
\caption{The main path of the rp-process in a one-zone X-ray burst model
\protect\cite{SAB01}. Stable nuclides are grey, nuclides with masses 
experimentally known to better
than 10~keV are red, other nuclides with experimentally known masses 
are orange, nuclides with experimentally known half-lives but 
unknown masses are green, and nuclei that have only been 
identified in experiments are blue. \label{FigRp}}
\end{figure}

As shown in Fig.~\ref{FigRp} 
\cite{SAG97} the rp process in X-ray bursts follows a path away from
the valley of stability through $\beta$-unstable nuclei. 
It is dominated
by (p,$\gamma$) reactions and $\beta$-decays and extends up to the mass $\sim$64-106
region depending on the astrophysical conditions. 
For lighter nuclei (up to mass $\sim$40) the ($\alpha$,p) reaction can
compete successfully with proton capture at X-ray burst temperatures.
The ($\alpha$,p) reaction followed by a radiative proton capture
(p,$\gamma$) is a very effective way for the production of  heavier
nuclei without waiting for an intermediate beta decay. This so-called
$\alpha$p process is thought to determine the early phase of an X-ray burst
lightcurve when the increasing temperature enables ($\alpha$,p) reactions
at successively heavier nuclei. Depending on the peak temperature
reached in a particular burst the final $\alpha$p-process can 
extend up to Sc, mainly through the sequence $^{14}$O($\alpha$,p)$^{17}$F(p,$\gamma$)$^{18}$Ne
($\alpha$,p)$^{21}$Na(p,$\gamma$)$^{22}$Mg($\alpha$,p)$^{25}$Al(p,$\gamma$)
$^{26}$Si($\alpha$,p)$^{29}$P(p,$\gamma$)$^{30}$S($\alpha$,p)$^{33}$Cl
(p,$\gamma$)$^{34}$Ar($\alpha$,p)$^{37}$K(p,$\gamma$)$^{38}$Ca($\alpha$,p)$^{21}$Sc. 
At several points along the path short-lived $\beta$-decays can compete leading 
to several parallel flows. 

Most of the nuclei probed in the rp-process do not exist as stable nuclei
but the majority of them have been produced in the
laboratory, as indicated in Fig.~\ref{FigRp}.  Besides 
particle stability half-lives and masses 
are the other crucial
input parameters for determining the reaction paths in the rp-process. 

\subsection{Nuclear Masses}

In the rp process 
proton capture is favored compared to $\beta$-decay until a nucleus
with a small (p,$\gamma$) Q value is reached where an equilibrium  between
(p,$\gamma$) capture and its inverse ($\gamma$,p) photodisintegration is established. At
this point the rp process has to wait until via $\beta^+$ decay a less
proton-rich nucleus is formed. Since the equilibrium depends exponentially
on the (p,$\gamma$) Q-value the mass of the waiting point nucleus itself as 
well the masses of the isotones which can be reached via 
one and two proton capture reactions 
have to be known with a precision of about $kT \approx 100$ keV
(with $k$ being the Boltzmann constant 
and T a typical temperature).

Nuclear masses are also important for the determination of proton
capture rates. For many of these rates along the rp-process path Q-values are
too high to establish a (p,$\gamma$)-($\gamma$,p) equilibrium 
at all times, but they are still low enough for a few individual 
resonances to dominate the reaction rate. Because of the 
exponential dependence of the proton capture rate on the resonance
energy, masses (and excitation energies) need to be known to better than 
10~keV in those cases.
As Fig.~\ref{FigRp} shows, for most of the rp-process, 
especially beyond Ti, masses are 
still not accurate enough for reliable reaction rate calculations or
the planning of direct reaction rate measurements with radioactive beams. 

For the most part masses in the rp-process can be theoretically 
predicted using extrapolations \cite{AME03} or, beyond the $N=Z$ line,
Coulomb shift calculations \cite{AB1}. In some cases where the mass or the 
mirror nucleus
is known with high accuracy, the latter method can reach 100~keV accuracy. 
While this represents a considerable improvement over global mass models
\cite{SAG97}, 
the uncertainties in theoretical mass predictions are still too large
for astrophysical applications. Experimental data are therefore needed. 

Considerable progress has been made during the past years at various
laboratories to measure masses of rp process nuclei (see Fig.~\ref{FigRp}),
but much remains to be done \cite{LUN03}. As Fig.~\ref{FigRp} shows, already beyond
magnesium not all reaction Q-values are known to better than 10~keV,
and beyond vanadium hardly any are sufficiently accurate. 
One new technique are mass measurements of unstable nuclei in the ESR 
storage ring at GSI using an isochroneous tune and a time-of-flight 
detector \cite{GSI1,GSI2}. The advantage of this technique is the low
half-life limit of a few microseconds that includes all particle 
bound nuclei. Conventional $\beta$-endpoint \cite{ND68SE} and time-of-flight
techniques \cite{TOF1} have also been used. 
A major step forward in this field, however, 
was the use of Penning traps 
with radioactive proton rich beams. These systems can achieve the needed
accuracies of 10-100~keV or better for even the heaviest rp-process
nuclei.  Several of these mass spectrometers  are now
operational worldwide \cite{PENN1,PENN2,PENN3,SBL03}. 
The nuclei are produced using various nuclear
reactions, they are thermalized, cooled and then injected into the
Penning trap. For a mass measurement the isotope has to live at least
100 ms, a condition which is fulfilled for practically all rp-process
nuclei.\\

Along the proton dripline the
pairing force can lead to the
stability of certain even-Z nuclei while the neighboring Z+1 nuclei are
proton unstable. Examples are the N=34 and 36 
N=Z nuclei
$^{68}$Se and $^{72}$Kr
which have relatively long half-lives of 35.5 s and 17 s, respectively, while
the isotones $^{69}$Br and $^{73}$Rb are particle-unbound. For the $^{68}$Se
case the $\beta$-decay
waiting point can be bridged through the successive capture of two protons
via the particle-unstable $^{69}$Br leading to
the particle-stable nucleus $^{70}$Kr. The mass of the
particle-unstable $^{69}$Br can only be measured through transfer reactions
using radioactive beams.
However, an estimate for the masses of these unstable nuclei can be obtained
using Coulomb displacement energies from mirror nuclei \cite{AB1}.  \\
A recent mass measurement of $^{68}$Se\cite{CLA04} together with 
estimates of the mass
of $^{69}$Br has led to the conclusion that $^{68}$Se remains a waiting
point in the rp process. Similar results have been obtained for 
$^{64}$Ge \cite{GE64}
and $^{72}$Kr \cite{KR72}. In the latter case the masses of $^{72-74}$Kr had been
measured providing together with Coulomb shift calculations
good estimates for the masses of the entire relevant isotone chain $^{72}$Kr, $^{73}$Rb,
and $^{74}$Kr. As in the case of $^{68}$Se it was found that $^{72}$Kr probably remains a
significant waiting point in the rp-process, but it was also pointed out
that because of uncertainties in the proton capture rates final
conclusions cannot be drawn yet.

In summary, for most of the important waiting point nuclei
such as $^{64}$Ge, $^{68}$Se, $^{72}$Kr and $^{104}$Sn,
masses and half lives are now known with sufficient accuracy. However,
what is needed for rp-process calculations are proton capture Q-values. For
$^{64}$Ge, $^{68}$Se, and $^{72}$Kr the mass of the proton
capture isotone is not yet known experimentally. For $^{104}$Sn
the proton capture Q-value has been determined from the detection of
proton decay in $^{105}$Sb \cite{SB1051,SB1052}. However, the two experiments give
somewhat conflicting results of either 478$\pm$15~keV or 550$\pm$30~keV  for the 
proton energy. 
Therefore, despite 
the considerable progress that has been made, accurate
experimental proton capture Q-values are still not available for the 
major rp-process waiting points.\\

\subsection{$\beta$-decay half-lives}
$\beta$-decay half-lives are critical for rp-process calculations
as they determine the timescale for the buildup of heavier elements
and energy generation as well as 
the final abundance distribution. 
As Fig.~\ref{FigRp} shows, most of the important $\beta$-decay half-lives in 
the rp-process have been measured in radioactive beam experiments with a
few exceptions such as $^{95,96}$Cd. In principle, theoretical corrections need to 
be applied to the experimental half-lives to account for the actual 
astrophysical environment during an X-ray burst. This has been 
investigated by Fuller, Fowler and Newman 1982 \cite{FFN82} and
more recently by Pruet et al. 2003 \cite{PrF03}, who tabulate 
corrected weak decay rates as a function of temperature and density
for use in X-ray burst calculations. 
At temperatures
of up to 2~GK low-lying states can be thermally populated, and the decay
rates from these states can differ from the ground state decay rates
measured in experiments. As the most important waiting point nuclei in 
the rp-process tend to be even-even nuclei with relatively high-lying 
first excited states this is usually not a large effect \cite{SAG97}. 
Exceptions might occur around $^{80}$Zr where a large ground state
deformation lowers the energy of the first excited state considerably. 
The second correction is needed to take electron capture into account 
properly. Nuclei in the X-ray burst environment are fully ionized and
bound state electron capture that might be present in 
the laboratory, is not important in the stellar environment. On the 
other hand, electron densities during X-ray bursts can be sufficiently 
large for continuum electron capture to play a role. However, 
as the majority of the important rp-process waiting points are 
far from stability and have large $\beta$-decay Q-values 
electron capture in the laboratory and in the star is in most cases negligible
at the 10\% level. In that case the use of measured terrestrial half-lives
in rp-process calculations is appropriate. However, such corrections can become important 
closer to stability, for example at the onset of the rp-process
below Zn and in particular X-ray bursts where the rp-process ends in the $A=60-72$
mass range and moves closer to stability during the final cooling phase. 

$\beta$-decays in the rp-process are not only important bottle-necks
in the reaction flow, but serve also as the major nuclear energy source owing to their
large positive Q-values. To accurately determine the energy production 
of the rp-process, the actual 
$\beta$-decay scheme needs to be known to determine the amount of 
energy that is lost via neutrino emission. While most of the $\beta$-decay
half-lives are measured, in many cases the actual $\beta$-strength functions
have not been determined experimentally. Therefore theoretical
$\beta$-decay calculations \cite{FFN82,PrF03} have to be used to determine 
the neutrino energy loss during X-ray bursts. \\

\subsection{Astrophysical Reaction Rates.}
Because the majority of nuclei involved in the rp process are unstable most of the 
reaction rates so far had to be estimated based on
theoretical grounds. While a statistical Hauser-Feshbach treatment is possible for
nuclei with high (p,$\gamma$) Q-values where states at high excitation 
energy are being populated\cite{RAU1}, in many cases the individual level structure 
of a particular nucleus can dominate the reaction rate. Shell model 
calculations have been performed for these cases in the sd- \cite{SD1} and 
fp-shell \cite{FP1} and these rates are used widely in rp-process
calculations. However, the accuracy of the predicted excitation energies
of individual levels is at most around 100~keV. For resonant 
reaction rates this translates into uncertainties of many orders of magnitude.
Even in cases where the level energies have 
been determined experimentally, an uncertainty of at least a factor of 2-3
remains. 
For this reason 
direct measurements of critical reactions along the rp-process path are 
urgently needed. 

Reaction rates in the rp-process need to be determined for a wide range of 
temperatures up to about 2~GK. While X-ray bursts ignite at temperatures
around 0.2~GK, the lower temperature limit for the reaction rate determination
is usually set by the condition that the reaction rate timescale has to be less
than the typical X-ray burst timescale of $\sim$10~s.
For typical X-ray burst conditions with 
hydrogen mass fractions of 0.5 and mass densities of 10$^6$~g/cm$^3$  one 
finds that a proton induced reaction rate is important once it exceeds
2$\times 10^{-7}$ cm$^3$/s/mole. As an example, proton 
capture rates on Ne, Cl, and Ge need to be known for temperatures
above 0.2, 0.3, and 0.5 ~GK respectively. 

With the exception of a few long-lived nuclei (e.g. $^{18}$F, $^{22}$Na,
$^{26}$Al, $^{44}$Ti and $^{56}$Ni) which could be produced as targets
in standard kinematics experiments using proton or alpha beams all other
isotopes along the rp path have very short half-lives and require the use
of the so-called inverse kinematics, where a proton or He target is
bombarded with a heavier radioactive beam.  Some of these beams have become
available during the last 15 years at several laboratories, leading to a
new era in nuclear astrophysics. Since the beam intensities at these
first generation facilities are still 3-5 orders of magnitude below
the intensities available with stable beams new, high efficiency
detector systems optimized for reaction studies in inverse kinematics
had to be developed. So far only a handful of experimental (p,$\alpha$) and
($\alpha$,p) reaction rates have been measured directly during the past 
few years. The direct measurement of a (p,$\gamma$) reaction with 
a radioactive beam has only succeeded in two cases, 
the $^{13}$N(p,$\gamma$)$^{14}$O reaction at Louvain-la-Neuve
\cite{DEC91,DEL93}
and the $^{21}$Na(p,$\gamma$)$^{22}$Mg reaction at TRIUMF
\cite{BIS03}. The rates of these reactions are discussed below.\\

For the many cases where a direct measurement of the reaction rate is still 
outside the experimental capabilities, indirect approaches can be used.
They are important when resonances are too weak, when the exact location 
of the resonances are not known or when the radioactive beams needed for a 
direct measurement can not yet be produced. They are also needed to determine
capture rates on low-lying thermally excited states in the target, which 
at the extreme temperatures reached in X-ray bursts can be important in some cases. 
For example, elastic scattering 
can be used to locate resonances or determine their total widths, and transfer 
reactions can give information about partial widths and decay properties. 
Gamma detector arrays, such as GAMMASPHERE \cite{Lee90} or SEGA \cite{Mue01} with their excellent energy 
resolution can also be used to identify states that dominate the reaction 
rate for nuclei away from stability.\\ 

In some cases it is more convenient to measure the time-inverse of the
reaction that actually occurs in the astrophysical environment. The
information on the forward reaction is then obtained through the principle
of detailed balance. Coulomb dissociation is the time-inverse of a direct
capture reaction and has been successfully used in some (mainly lighter)
systems, e.g. $^{13}$N(p,$\gamma$)$^{14}$O\cite{MOT91,KIE93}. The technique can give an
increase in yield by a factor of 10$^5$ over the forward capture reaction
and can thus be used with relatively weak radioactive beams. It probes,
however,  only the $\gamma$-transition to the ground state and there are 
additional difficulties with transitions of mixed multipolarity.\\

Other examples of the successful use of time-inverse reactions which
are of importance to the study of X-ray bursts are studies of ($\alpha$,p)
reactions through their time-inverse partner (p,$\alpha$), thereby avoiding 
the use of a He target. A drawback of this technique is that contributions 
from excited states in the final nucleus have to be determined separately, 
e.g. through a measurement of elastic and inelastic scattering.

In the following the results from recent measurements of reaction 
rates in X-ray bursts with
radioactive beams or targets using both, direct and indirect techniques
are briefly discussed.\\

\subsubsection{The triple-$\alpha$ reaction}

The rate for the triple-$\alpha$ reaction which triggers the X-ray bursts
and serves as the starting point and major initial bottle-neck of the rp-processs 
(see Fig.1) has recently seen some changes resulting from the
spin and energy determination of a resonance at E$_x \sim 11$~MeV\cite{fyn2005}. The main 
modifications are an increase in the rate at low temperatures 
(below 5$\times$10$^7$ K) and a steep decrease by more than an order 
of magnitude at temperatures above 2.5$\times$10$^9$ K. More 
calculations need to be done to investigate the effect of the changes 
in the reaction rate on the rp process. 

\subsubsection{The $^{13}$N(p,$\gamma$)$^{14}$O Reaction}

The $^{13}$N(p,$\gamma$)$^{14}$O reaction can bypass the slow $\beta$-decay
of $^{13}$N (T$_{1/2}$=9.96m) leading to the so-called hot CNO cycle where
the energy production is independent of the temperature and limited by
the beta decays of $^{14}$O(T$_{1/2}$=70.6 s) and $^{15}$O(T$_{1/2}$=2.03 m).
This reaction was measured directly in a pioneering experiment with a
$^{13}$N beam at Louvain-la-Neuve\cite{DEC91,DEL93} as well as with indirect (Coulomb
dissociation) techniques at RIKEN\cite{MOT91} and GANIL\cite{KIE93}. 
The results from these
experiments agree among each other 
giving a width $\Gamma_{\gamma}$ of 3.0$\pm$0.5 eV and a resonance
strength of 9$\pm$1.5 eV. A recent measurement extracting the Asymptotic
Normalization Coefficient (ANC) from a transfer reaction measurement has led
to a slight modification of the reaction rate which agrees with the
tabulated values \cite{NAC03} at temperatures above T$_9\sim$0.5 but
exhibits an increase by a factor of 1.5-2 below T$_9\sim$0.25\cite{TAN04}.

\subsubsection{The $^{14}$O($\alpha$,p)$^{17}$F Reaction.} 

This reaction has been studied through its time inverse
$^{17}$F(p,$\alpha$)$^{14}$O reaction at ANL\cite{Har99} and ORNL\cite{BLA01}. The transitions to
the first excited 1/2$^+$ state in $^{17}$F was addressed through independent
measurements of the $^{17}$F(p,p')$^{17}$F reaction\cite{HAR02,BLA03}. The cross sections in
these two experiments were found to agree within their experimental
uncertainties. In a more recent experiment this reaction was measured
directly with a $^{14}$O beam and a cryogenic He target\cite{NOT04}. There is some 
disagreement between the direct and indirect measurements that need to 
be addressed in future measurements.\\

\subsubsection{The $^{17}$F(p,$\gamma$)$^{18}$Ne Reaction.}

This reaction rate is part of the ($\alpha$,p) process following the
$^{14}$O($\alpha$,p)$^{17}$F reaction. It depends critically
on the exact location of a 3$^+$ state which is expected from mirror symmetry
arguments at an energy of 4.5 MeV. This state was first unambiguously
identified in experiments with $^{17}$F beams at ORNL\cite{BAR99,BAR00,GAL00}. From the resonant
strength whose value still depends somewhat on properties of the mirror
nucleus one finds that the  reaction rate for temperatures above 
T$_{9}\sim$0.5 
is determined by this state. \\

\subsubsection{The $^{18}$Ne($\alpha$,p)$^{21}$Na reaction.} 

This reaction was studied directly with a $^{18}$Ne beam and an active He
target\cite{BRA99,GRO02} at excitation energies which, however, are too high
to be relevant for X-ray bursts. Furthermore, the cross sections obtained in
these experiments seem to exhaust the Wigner limits for states at an
excitation energy of ~11 MeV. Recent experiments using the indirect
time-inverse $^{21}$Na(p,$\alpha$)$^{18}$Ne approach have found considerably
smaller yields\cite{SIN04}.  These discrepancies require further studies
and experiments at several radioactive beam facilities are planned for the
future. \\

\subsubsection{The $^{19}$Ne(p,$\gamma$)$^{20}$Na Reaction.}

As part of the $^{15}$O($\alpha,\gamma$)$^{19}$Ne(p,$\gamma$)$^{20}$Na chain
this reaction is at the onset of the rp process. A direct study using a
$^{19}$Ne beam and detecting the delayed decay products from $^{20}$Na 
\cite{PAG94,VAN98} has so far only given an upper limit of the
first state above the proton 
threshold (see also \cite{CAC04}. However, even the spin of this state is still being
debated\cite{LAM90,KUB92,BRO93,SEW04} and, 
thus, the value of the reaction rate for this reaction is still uncertain 
by orders of magnitude.\\

\subsubsection{The $^{21}$Na(p,$\gamma$)$^{22}$Mg Reaction.}

This reaction is again part of the ($\alpha$,p) process, following the
$^{18}$Ne($\alpha$,p)$^{21}$Na reaction. 
The resonant strengths
of seven resonances in $^{22}$Mg have been directly measured with a
$^{21}$Na beam at TRIUMF\cite{AZU03,DAU04}. \\

\subsubsection{The $^{22}$Na(p,$\gamma$)$^{23}$Mg Reaction.}

This reaction has been studied by direct and indirect techniques.
Bombarding a radioactive $^{22}$Na (T$_{1/2}$=2.6 y) target with a proton 
beam in two experiments has led to a reaction rate which is smaller by 
about an order of magnitude when compared to a pure theoretical
estimate\cite{SEU90,STE96}. 
More recent indirect measurements using GAMMASPHERE resulted in an 
increase of this rate due to the addition of new resonances which were 
not included in the previous analyses\cite{JEN04}. \\

\subsubsection{The $^{22}$Mg(p,$\gamma$)$^{23}$Al Reaction}

$^{22}$Mg is a potential local waiting point where proton capture competes
with the ($\alpha$,p) reaction. A recent measurement of the
$^{24}$Mg($^7$Li,$^8$He)$^{23}$Al reaction established the mass and the
energy of the first excited state in $^{23}$Al\cite{CAG01}. The astrophysical
reaction
rate for the $^{22}$Mg(p,$\gamma$) reaction was then obtained using a
spectroscopic factor based on shell model calculations. A recent measurement
of Coulomb dissociation of $^{23}$Al\cite{GOM04} resulted in a resonance
strength which is in good agreement with the value used in
Ref.\cite{CAG01}. These measurements have reduced the uncertainties of
the (p,$\gamma$) rate considerably. However, because of its small Q-value the
$^{22}$Mg(p,$\gamma$)$^{23}$Al reaction is in equilibrium with its inverse
reaction $^{23}$Al($\gamma$,p)$^{22}$Mg and, thus, very little $^{23}$Al is
produced.
The influence of the 2-proton capture reaction on $^{22}$Mg forming $^{24}$Si
which becomes important at higher temperatures has been discussed in
Ref.\cite{SCH97}.

\subsubsection{The $^{32}$Cl(p,$\gamma$)$^{33}$Ar Reaction.}

The reaction rate used in previous calculations was entirely based on
shell-model calculations with uncertainties in the excitation energy
around 100 keV. Populating the states in $^{33}$Ar above the proton threshold
via the $^{34}$Ar(p,d)$^{33}$Ar reaction and studying their $\gamma$ decay 
has allowed to put much smaller error bars on the excitation energies 
of some of the critical levels\cite{CLE04}. This has reduced the uncertainty in the 
reaction rate from 3 orders of magnitude to a factor of 3 at the 
critical temperatures around 0.3~GK and to a factor of 6 at the 
highest X-ray burst temperatures.  For high temperatures
additional states, which have not been identified so far 
might contribute as well. Further studies for this system are clearly 
desirable.\\

\subsubsection{The $^{56}$Ni(p,$\gamma$)$^{57}$Cu Reaction.}

The sequence of ($\alpha$,p) and (p,$\gamma$) reactions leading to the
formation of heavier nuclei in a X-ray bursts proceeds mainly through the
nucleus $^{56}$Ni which is a waiting point in the reaction flow. Because
of the closed-shell nature of $^{56}$Ni the Q value for the 
$^{56}$Ni(p,$\gamma$)$^{57}$Cu reaction is quite
small (Q=0.695 MeV) so that the cross section can not be
reliably estimated
using statistical model calculations. The rate is dominated by transitions to 
the first two excited states in $^{57}$Cu. A direct measurement of the
cross section is impossible using present-day capabilities and
thus indirect methods have to be employed. In such an approach the
one-particle transfer reactions $^{56}$Ni(d,p)$^{57}$Ni\cite{REH98,REH00} and
$^{56}$Ni($^3$He,d)$^{57}$Cu\cite{JIA99} have been measured with a weak (3x10$^4$/s)
$^{56}$Ni beam. Using the spectroscopic factors from these measurements,
calculated proton penetrabilities, and
$\gamma$-widths from the $^{57}$Ni mirror \cite{ZHO96} 
an estimate for the
$^{56}$Ni(p,$\gamma$)$^{57}$Cu reaction rate could be determined which 
was found to be 10 times larger than previous estimates\cite{Van94,ZHO96}.
For temperatures above 1~GK the reaction rate was later slightly 
revised due to an improved treatment of the $\gamma$-widths 
\cite{For01}.\\

In addition to radioactive beam mesurements, experiments
with stable beams
continue to play an important role for the lower part of the
$\alpha$p- and rp-process reaction chains below about Ca. Recent examples
of experiments that use transfer reactions with stable beams to
reach nuclei in the rp-process include measurements using
the $^{20}$Ne($^3$He,$\alpha$)$^{19}$Ne reaction at ANL\cite{REH03},
the $^{21}$Ne(p,t)$^{19}$Ne reaction at KVI \cite{DAV1,DAV2},
the $^{24}$Mg($^4$He,$^6$He)$^{22}$Mg reaction at RCNP \cite{MG22},
the $^{25}$Mg($^3$He,$^6$He)$^{22}$Mg reaction at Yale \cite{CAG02},
the $^{24}$Mg(p,t)$^{22}$Mg reaction at CNS Tokyo\cite{BAT01},
the $^{12}$C($^{12}$C,2n)$^{22}$Mg reaction at ANL\cite{SEW05},
the $^{24}$Mg($^7$Li,$^8$He)$^{23}$Al and
$^{28}$Si($^7$Li,$^8$He)$^{27}$P\cite{CAG01}
reactions at MSU and a variety of reactions to access states in
$^{19}$Ne,$^{26}$Si and $^{26}$Al at ANL, Notre Dame, Princeton, 
and Yale. 

From the compilation of the experimental reaction rates
given above it can be seen that only a small percentage of the astrophysical
reaction rates which are relevant for X-ray bursts are based on experimental
data. For most of them only theoretical estimates are available. The
same holds for many electron capture rates and pycnonuclear fusion cross
sections involving neutron-rich nuclei. While significant progress is 
expected with existing facilities, a new generation of radioactive 
beam facilities such as RIA is needed to address all the nuclear 
physics questions related to accreting neutron stars in X-ray 
binaries. 

The authors thank Ed Brown for useful discussions concerning this 
manuscript. This work has been supported by the Joint Institute 
for Nuclear Astrophysics (JINA) under NSF grant PHY 02-16783. H.S. acknowledges
support from NSF grant PHY 0110253. K.E.R is supported by the U.S. 
Department of Energy, Office
of Nuclear Physics under contract W-31-109-ENG-38.


\begin{thebibliography}{999999}

\bibitem{Psa04}
D. Psaltis, astro-ph/0410536 (2004).

\bibitem{StB03}
T.E. Strohmayer and L. Bildsten, Compact Stellar X-ray Sources, ed.
W. H. G. Lewin and M. van der Klies (Cambridge: Cambridge Univ. Press),
astro-ph/0301544 (2003). 

%===================== Lines ===============================================

\bibitem{BCP03}
L. Bildsten, P. Chang, and F. Paerels, Ap. J. 591 (2003), L29.

\bibitem{CPM02}
J. Cottam, F. Paerels, and M. Mendez, Nature 420 (2002) 51.

\bibitem{BSW92}
L. Bildsten, E. E. Salpeter, and I. Wasserman Ap. J. 384 (1992) 143.

\bibitem{StV04}
T. E. Strohmayer, and A. R. Villarreal Ap. J. 614 (2004) 121.

\bibitem{Jos77}
P. C. Joss, Nature 270 (1977) 310. 

\bibitem{Nevin06}
N. N. Weinberg, L. Bildsten, and H. Schatz, Ap. J. 639 (2006) 1018.

%================= XRB ====================================================
\bibitem{Gri76}
J. Grindlay et al. Ap. J. 205 (1976) L127.

\bibitem{BCE76}
R. D. Belian, J. P. Conner, and W. D. Evans, Ap. J. 206 (1976) L135.

\bibitem{WoT76}
S.~E. Woosley and R.~E. Taam, Nature 263  101  (1976).

\bibitem{MaC77}
L. Maraschi and A. Cavaliere, Highlights of Astronomy 4 (1977) 127.

\bibitem{LaL78}
D. Q. Lamb and F. K. Lamb, Ap. J., 220 (1978) 220. 

\bibitem{HaV75}
C. J. Hansen and H. M. Van Horn, Ap. J. 195 (1975) 735.

\bibitem{Bil97}
L. Bildsten,  in {\em The Many Faces of Neutron Stars}, edited by A. Alpar, L.
  Bucceri, and J. Van~Paradijs (Dordrecht, Kluwer, 1998), pp.\
  astro--ph/9709094.

\bibitem{LPT93}
W.~H.~G. Lewin, J. van Paradijs, and R.~E. Taam, Space Sci. Rev. 62 (1993)  233.

\bibitem{Str96}
T. E. Strohmayer et al. Ap. J. 469 (1996) L9. 

\bibitem{CMB02}
A. Cumming et al. Ap. J. 564 (2002) 343.

\bibitem{PiB05}
A. L. Piro and L. Bildsten, submitted to Ap. J. (2005), astro-ph/0502546.

\bibitem{CuB98}
A. Cumming and L. Bildsten, Ap. J. 506 (1998) 842.

\bibitem{MVH88}
P. N. McDermott, H. M. van Horn, and C. J. Hansen, Ap. J. 325 (1988) 725.

\bibitem{CMM03}
D. Chakrabarty et al. Nature 424 (2003) 42.

\bibitem{SMS03}
T. E. Strohmayer et al. Ap. J. 596 (2003) L67.

\bibitem{MGC04}
M. P. Muno, D. K. Galloway, and D. Chakrabarty, Ap. J. 608, (2004), 930.

\bibitem{Str04}
T. E. Strohmayer, astro-ph/0401465 (2004). 

\bibitem{BSC04}
S. Bhattacharyya et al., astro-ph/0402534 (2004).

\bibitem{WHC04}
S. E. Woosley et al., Ap. J. Suppl. 151 (2004) 75.

\bibitem{WaW81}
R.~K. Wallace and S.~E. Woosley, Ap. J. Suppl. 45 (1981)  389.

\bibitem{Van94}
  L. van Wormer et al. Ap. J. 432 (1994) 326.

\bibitem{SAG97}
H. Schatz {\it et~al.}, Phys. Rep. 294 (1998) 167.

\bibitem{SAB01}
H. Schatz {\it et~al.}, Phys. Rev. Lett. 86 (2001)  3471.

\bibitem{WiS99}
M. Wiescher and H. Schatz, Journ. Phys. G Topical Review 25 (1999)  R133.

\bibitem{BHM98}
R.~N. Boyd, M. Hencheck, and B.~S. Meyer,  in {\em International Symposium on
  Origin of Matter and Evolution of Galaxies 97, Atami, Japan}, edited by S.
  Kubono, T. Kajino, K.~I. Nomoto, and I. Tanihata (World Scientific, New
  Jersey, Singapore, 1998), p.\ 350.

\bibitem{TWL96}
R.~E. Taam, S.~E. Woosley, and D.~Q. Lamb, Ap. J. 459 (1996) 271.

\bibitem{KHA99}
O. Koike  {\it et~al.}, Astron. Astrophys. 342 (1999) 464.

\bibitem{SBC98}
H. Schatz  {\it et~al.}, Ap. J. 524 (1999) 1014.

\bibitem{KHK04}
O. Koike et al. Ap. J. 603 (2004) 242. 

\bibitem{TBF01}
F.-K. Thielemann et~al., Prog. in Part. and Nucl. Phys. 46 (2001) 46.

\bibitem{FTW04}
J. L. Fisker, F.-K. Thielemann, and M. Wiescher, Ap. J. 608 (2004) 61. 

\bibitem{CCG87}
M. Czerny, B. Czerny, and J. E. Grindlay, Ap. J. 312 (1987) 122.

\bibitem{FTW92}
I. Fushiki et al. Ap. J. 390 (1992) 634. 

\bibitem{Cum03}
A. Cumming, astro-ph/0309626 (2003).

\bibitem{Tam80}
R. E. Taam, Ap. J. 241, (1980), 358

\bibitem{GCK04}
D. K. Galloway et al. Ap. J. 601 (2004) 466.

\bibitem{FrW82}
B. A. Fryxell and S. E. Woosley, Ap. J. 261 (1982) 332.

\bibitem{Bil95}
L. Bildsten, Ap. J. 438 (1995) 852. 

\bibitem{SLU02}
A. Spitkovsky, Y. Levin, and G. Ushomirsky, Ap. J. 566 (2002) 1018.

\bibitem{Zin02}
M. Zingale et al. astro-ph/0211336 (2002).

\bibitem{Zin01}
M. Zingale et. al. Ap. J. Suppl. 133 (2001) 195.

\bibitem{Cor03}
R. Cornelisse et al. A\&A 405 (2003), 1033.

\bibitem{NaH03}
R. Narayan and J. S. Heyl, Ap. J. 599 (2003) 419.

\bibitem{Heg04}
A. Heger, private communication (2004). 

\bibitem{HBM95}
M. Hencheck et al. Nuclei in the Cosmos III, Third International Symposium on Nuclear Astrophysics, Proceedings of the Symposium held at the National Laboratories of Gran Sasso, Assergi, L'Aquil, Italy, July 1994. New York: American Institute of Physics Press. Edited by Maurizio Busso, Claudia M. Raiteri, and Roberto Gallino. AIP Conference Proceedings, Vol. 327, 1995, p.331.

\bibitem{LAM92}
D.~L. Lambert, Astron. Astrophys. Rev. 3 (1992) 201.

\bibitem{ArG03}
M. Arnould and S. Goriely, Phys. Rep. 384 (2003) 1.

\bibitem{CRZ00}
V. Costa  {\it et~al.}, Astron. Astrophys. 358 (2000) L67.

\bibitem{JKM01}
M. Jaeger et al. Phys. Rev. Lett. 87 (2001) 250.

\bibitem{GJH02}
S. Goriely et al. Astron. Astrophys. 383 (2002) L27. 

\bibitem{DRM02}
N. Dauphas et al. astro-ph/0211452 (2002).

\bibitem{ICC04}
J. M. M. in't Zandt, R. Cornelisse, and A. Cumming, Astron. Astrophys. 426
(2004) 257.

\bibitem{ICK04}
J. M. M. in't Zandt et al. astro-ph/0407087 (2004).

\bibitem{StM02}
T. E. Strohmayer and C. B. Markwardt, Ap. J. 577 (2002) 337.

\bibitem{StB02}
T. E. Strohmayer and E. F. Brown, Ap. J. 566 (2002) 1042.

\bibitem{CuB01}
A. Cumming and L. Bildsten, Ap. J. 559 (2001) L127.

\bibitem{SBC03}
H. Schatz, L. Bildsten, and A. Cumming, Ap. J. 583 (2003) L90.

\bibitem{CuM04}
A. Cumming and J. Macbeth, Ap. J. 603 (2004) L37.

\bibitem{Bro04}
E. F. Brown, Ap. J. 614 (2004) 57.

\bibitem{CSL97}
W. Chen, C. R. Shrader, and M. Livio, Ap. J. 491 (1997) 312.

\bibitem{CCM98}
S. Campana et al. A\&A rev. 8 (1998) 279.

\bibitem{CGP01}
M. Colpi, U. Geppert, D. Page, and A. Possenti, Ap. J. {\bf 548},  L175
  (2001).

\bibitem{YLH03}
D. G. Yakolev, K. P. levenfish, and P. Haensel, Astron. Astrophys. 407
(2004) 265.

\bibitem{BBR98}
E.~F. Brown, L. Bildsten, and R.~E. Rutledge, Ap. J. 504 (1998)  L95.

\bibitem{RBB00}
R. Rutledge {\it et~al.}, Ap. J. 529 (2000)  985.

\bibitem{RBB99}
R. E. Rutledge {\it et al.},Ap. J.  514 (1999) 945. 

\bibitem{RBB02}
R. E. Rutledge et al. Ap. J. 577 (2002) 346.

\bibitem{Wij04}
R. Wjinands, astro-ph/0405089 (2004).

\bibitem{Wij04a}
R. Wijnands et al. Ap. J. 606 (2004) L61.

%================= Crust ====================================================

\bibitem{UGK98}
V. Urpin, U. Geppert, and D. Konenkov, M.N.R.A.S. 295 (1998) 907. 

\bibitem{BrB98}
E. F. Brown and L. Bildsten, Ap. J. 496 (1998) 915. 

\bibitem{CAZ04}
A. Cumming, P. Arras, and E. Zweibel, Ap. J. 609 (2004) 999.

\bibitem{HaZ90}
P. Haensel and J. L. Zdunik, Astron. Astrophys. 227 (1990) 431.

\bibitem{HaZ90a}
P. Haensel and J. L. Zdunik, Astron. Astrophys. 229 (1990) 117.

\bibitem{HaZ03}
P. Haensel and J. L. Zdunik, Astron. Astrophys. 404 (2003) 33.

\bibitem{Bil98}
L. Bildsten Ap. J. Lett. 501 (1998) 89.

\bibitem{UCB00}
G. Ushomirsky, C. Cutler, and L. Bildsten, Monthly Not. of the Royal Astron. Soc. 319 (2000) 902.

\bibitem{Jon04}
P. B. Jones, astro-ph/0403400 to appear in Phys. Rev. Lett. (2004). 
%%---------------------------nuclear reactions-------------------------------

\bibitem{ChW92}
A. E. Champagne and M. Wiescher, Ann. Rev. Nucl. Part. Sci. 42 (1992) 39.

\bibitem{WSC98}
M. Wiescher, H. Schatz, and A. E. Champagne, Phil. Trans. Royal. Soc. London 356 (1998) 2105.

\bibitem{GSI1} 
T. Radon et al. Phys. Rev. Lett. 78 (1997) 4701

\bibitem{GSI2}
J. Stadlmann et al., Phys. Lett. B586 (2004) 27

\bibitem{ND68SE} 
A. W\"{o}hr et al., Nucl. Phys. A742 (2004) 349
  
\bibitem{TOF1}
  M. Chartier. et al., Nucl. Phys. A637 (1998) 3
  
\bibitem{PENN1}
  H. Stolzenberg et al., Phys. Rev. Lett. 65 (1990) 3104 

\bibitem{PENN2}
  G. Savard et al., Nucl. Phys. A626 (1997) 353

\bibitem{PENN3}
  V. S. Kolhinen et al., Nucl. Instr. Methods A528 (2004) 776

\bibitem{SBL03}
  S. Schwarz et al. Nucl. Instr. Meth. B 204 (2003) 507

\bibitem{AME03}
  G. Audi, A. H. Wapstra and C. Thibault, Nucl. Phys. A 729 (2003) 129.

\bibitem{AB1}
B. A. Brown, R. R. C. Clement, H. Schatz, A. Volya, W. A. Richter,
Phys. Rev. C65 (2002) 045802

\bibitem{LUN03}
D. Lunney, J. M. Pearson, and C. Thibault, Rev. Mod. Phys. 75 (2003) 1021.

\bibitem{CLA04}
  J. A. Clark et al., Phys. Rev. Lett. 92 (2004) 192501

\bibitem{GE64} J. A. Clark et al., Proceed. of the Fourth Internat. Conf. on
Exotic Nuclei and Atomic Masses, (2004) 59

\bibitem{KR72}
  D. Rodriguez et al. Phys. Rev. Lett. 93 (2004) 161104

\bibitem{SB1051}
  R. J. Tighe et al., Phys. Rev. C49 (1994) R2871

\bibitem{SB1052}
  J. Friese et al., in Proceedings of the XXIV Workshop on Gross Properties
of Nuclei, Hirschegg, Austria 1996, edited by H. Feldmeier, J. Knoll and W.
N\"{o}renberg (GSI Darmstadt 1996) p. 123 

\bibitem{FFN82}
  G.M. Fuller, W.A. Fowler and M. J. Newman, Astrop.J. 252 (1982) 715

\bibitem{PrF03} J. Pruet and G. M. Fuller, Astrop. J. Suppl. 149 (2003) 189
  
\bibitem{RAU1}
  T. Rauscher, F. K. Thielemann, At. Data Nucle. Data Tabl. 75 (2000) 1

\bibitem{SD1}
  H. Herndl et al. Phys. Rev. C 52 (1995) 1078.

\bibitem{FP1}
  J. L. Fisker et al. Atomic Data and Nucl. Data Tab. 79 (2001) 241.

\bibitem{fyn2005}
  H. O. U. Fynbo et al. Nature 433 (2005) 136

\bibitem{DEC91}
P.Decrock et al., Phys. Rev. Lett. 67 (1991) 808 

\bibitem{DEL93}
Th. Delbar et al., Phys. Rev. C48 (1993) 3088

\bibitem{BIS03}
S. Bishop et al., Phys. Rev. Lett. 90 (2003) 162501

\bibitem{Lee90}
I. Y. Lee, Nucl. Phys. A 520 (1990) 641c

\bibitem{Mue01}
W. F. Mueller et al., Nucl. Instr. Methods A 466 (2001) 492.

\bibitem{MOT91}
T. Motobayashi et al., Phys. Lett. B264 (1991) 259

\bibitem{KIE93}
J. Kiener et al., Nucl. Phys. A552 (1993) 66

\bibitem{NAC03}
  C. Angulo, NACRE-Nuclear Astrophysics Compilation of Reaction Rates,
  http://pntpm.ulb.ac.be/nacre/ (2003)

\bibitem{TAN04}
X. Tang et al., Phys. Rev. C69 (2004) 055807

\bibitem{Har99}
B. Harss et al., Phys. Rev. Lett. 82 (1999) 3964

\bibitem{BLA01}
J.C. Blackmon et al., Nucl  Phys. A688 (2001) 142

\bibitem{HAR02}
B. Harss et al., Phys. Rev. C65 (2002) 035803

\bibitem{BLA03}
J. C. Blackmon et al., Nucl. Phys. A718 (2003) 127c

\bibitem{NOT04}
M. Notani et al., Nucl. Phys. A746 (2004) 113c 

\bibitem{BAR99}
D. W. Bardayan et al., Phys. Rev. Lett. 83 (1999) 45

\bibitem{BAR00}
D. W. Bardayan et al., Phys. Rev. C62 (2000) 055804

\bibitem{GAL00}
A. Galindo-Uribarri et al. Nucl. Instr. Methods B172 (2000) 647

\bibitem{BRA99}
W. Bradfield-Smith et al., Phys. Rev. C59 (1999) 3402

\bibitem{GRO02}
D. Groombridge et al., Phys. Rev. C66 (2002) 055802

\bibitem{SIN04}
S. Sinha et al., Bull. Am. Phys. Soc. (2004)

\bibitem{PAG94}
R. D. Page et al., Phys. Rev. Lett. 73 (1994) 3066

\bibitem{VAN98}
G. Vancraeynest et al., Phys. Rev. 57 (1998) 2711

\bibitem{CAC04}
M. Couder et al. Phys. Rev. C 69 (2004) 022801

\bibitem{LAM90}
L. O. Lamm et al., Nucl. Phys. A510 (1990) 503

\bibitem{KUB92}
  S. Kubono et al., Phys. Rev. C46 (1992) 361
  
\bibitem{BRO93}
  B. A. Brown, A. E. Champagne, H. T. Fortune and R. Sherr,
  Phys. Rev. C48 (1993) 1456
  
\bibitem{SEW04}
D. Seweryniak et al., Phys. Lett. B590 (2004) 170


\bibitem{AZU03}
R. E. Azuma et al., Nucl. Phys. A718 (2003) 199c

\bibitem{DAU04}
  J. M. D'Auria et al. Phys. Rev. C69 (2004) 065803
  
\bibitem{SEU90}
  S. Seuthe et al., Nucl. Phys. A514 (1990) 471
  
\bibitem{STE96}
  F. Stegm\"{u}ller et al., Nucl. Phys. A601, (1996) 168
  
\bibitem{JEN04}
  D. G. Jenkins et al., Phys. Rev. Lett.92 (2004) 031101

\bibitem{CAG01}
  J. A. Caggiano et al., Phys. Rev. C64 (2001) 025802
  
\bibitem{GOM04}
  T. Gomi et al., Nucl. Phys. A734 (2004) E77

\bibitem{SCH97}
  H. Schatz et al., Phys. Rev. Lett. 79 (1997) 3845
  
\bibitem{CLE04}
  R. R. C. Clement et al., Phys. Rev. Lett. 92 (2004) 172502

\bibitem{REH98}
  K. E. Rehm et al., Phys. Rev. Lett. 80 (1998) 676
  
\bibitem{REH00}
  K. E. Rehm et al., Nucl. Instr. Meth. A449 (2000) 208
  
\bibitem{JIA99}
  C. L. Jiang et al., Argonne National Laboratory Annual Report, ANL-99/12
  (1999) 11, unpublished

\bibitem{ZHO96}
  X. G. Zhou et al. Phys. Rev. C53 (1996) 982

\bibitem{For01}
  O. Forstner et al. Phys. Rev. C64 (2001) 5801.
  
\bibitem{REH03}
  K. E. Rehm et al., Phys. Rev. C67 (2003) 065809
  
\bibitem{DAV1}
  B. Davids et al., Phys. Rev. C67 (2003) 012801
  
\bibitem{DAV2}
  B. Davids et al., Phys. Rev. C67 (2003) 065808
  
\bibitem{MG22}
G. P. A. Berg et al., Nucl. Phys. A718, (2003) 608c

\bibitem{CAG02}
  J. A. Caggiano et al., Phys. Rev. C66 (2002) 015804
  
\bibitem{BAT01}
  N. Bateman et al., Phys. Rev. C63 (2001) 035803
  
\bibitem{SEW05}
  D. Seweryniak et al., accepted in Phys. Rev. Lett.
  
\end{thebibliography}
\end{document}